\documentclass[graybox]{svmult}

\usepackage{mathptmx}       
\usepackage{helvet}         
\usepackage{courier}        
\usepackage{type1cm}        
\usepackage{makeidx}         
\usepackage{graphicx}        
\usepackage{multicol}        
\usepackage[bottom]{footmisc}

\makeindex             

\begin{document}

\title*{Modelling stellar convection and pulsation in multidimensions using the ANTARES code}
\titlerunning{Stellar convection and pulsation}
\author{Eva Mundprecht and Herbert J. Muthsam}
\institute{Eva Mundprecht \at Faculty of Mathematics, University of Vienna, \email{eva.mundprecht@univie.ac.at}
\and Herbert J. Muthsam \at Faculty of Mathematics, University of Vienna, \email{herbert.muthsam@univie.ac.at}}
%
%
\maketitle

\abstract*{The ANTARES code has been designed for simulation of astrophysical flows in a variety of situations, in particular in the context of stellar physics. Here, we describe extensions as necessary to model the interaction of pulsation and convection in classical pulsating stars. These extensions encomprise the introduction of a spherical grid, moveable in the radial direction, specific forms of grid-refinement and considerations regarding radiative transfer. We then present the basic parameters of the cepheid we study more closely. For that star we provide a short discussion of patterns of the $\mbox{H}+\mbox{He\hspace{1pt}I}$ and the $\mbox{He\hspace{1pt}II}$ convection zones and the interaction with pulsation  seen in the $pdV$ work or atmospheric structures.}

\abstract{The ANTARES code has been designed for simulation of astrophysical flows in a variety of situations, in particular in the context of stellar physics. Here, we describe extensions as necessary to model the interaction of pulsation and convection in classical pulsating stars. These extensions encomprise the introduction of a spherical grid, moveable in the radial direction, specific forms of grid-refinement and considerations regarding radiative transfer. We then present the basic parameters of the cepheid we study more closely. For that star we provide a short discussion of patterns of the $\mbox{H}+\mbox{He\hspace{1pt}I}$ and the $\mbox{He\hspace{1pt}II}$ convection zones and the interaction with pulsation  seen in the $pdV$ work or atmospheric structures.
}

\section{Introduction}
\label{sec:1}
Time-dependent, non-linear models of the classical pulsating variables (cepheids, RR Lyr) are around since decades. Nearly all those models have treated the issue in one spatial dimension (plus time). As is well known, for the stars near the red edge the inclusion of convection is mandatory. Given the basic 1D computational setting, this had to be done using simplyfied models for convection. For a detailed discussion of shortcomings of this approach and their consequences see \cite{buch_rev_97},\cite{buch_rev_09}.  

We therefore have adapted the ANTARES code, \cite{muth_10}, in order to comply with the needs for modelling such stars in multidimensions (2D presently). In section \ref{sec:2} we address a few technical issues, in section \ref{sec:3} we discuss a specific model.

\section{Technical Issues}
\label{sec:2}

The simulations were performed on a stretched, polar and moving grid. We discuss the radiative transfer equation (RTE) and restrictions on time stepping in turn.

%
%


\runinhead{The Radiative Transfer Equation.} Near the surface, nontrivial energy exchange between gas and radiation is included via the RTE with grey opacities, while diffusion approximation is used for the deeper layers. The RTE in 1D is solved along single rays via the short characteristics method of Kunasz and Auer \cite{kun_auer}. 
In 2D either 12 or 24 ray directions are chosen according to the angular quadrature formulae of type A4 or A6 of Carlson \cite{carlson}, and the directions in each quadrant are arranged in a triangular pattern.
For each ray the points of entrance and exit plus the corresponding distances are determined. Since the grid moves this has to be redone every step. The RTE 
\begin{equation}
\mu \frac{\partial I}{\partial \tau }=S-I 
\end{equation} is then solved along each ray. This procedure is repeated recursively since after the first step one gets just the intensity on a single new point.

\runinhead{Time Stepping.} The time step for our (explicit) method is never limited by the classical CFL condition but is due to radiation transport.
Using the diffusion approximation also near the surface would enforce prohibitively small time steps,
\begin{equation}
\Delta t_{diffusive}\propto \min \left( \frac{3~c_{p}}{16\kappa\sigma T^{3}}\left( \frac{\kappa\rho}{k}\right)^{2} \right) \sim \min \left( \frac{min(\Delta r_{i},\Delta y_{i})^{2}}{\chi}\right)\mbox{.} 
\end{equation}
Fortunately, the radiative transfer equation allows the time step to stay in contact to the  time scale for relaxing a temperature perturbation on the scale of the grid size by radiation (Spiegel's relation, \cite{spiegel}),
\begin{equation}
\Delta t_{rad}\propto \min \left( \frac{c_{p}}{16\kappa\sigma T^{3}}\left( 1-\frac{\kappa\rho}{k}arccot\frac{\kappa\rho}{k}\right)^{-1}\right)\mbox{,}
\end{equation}
with $k=\frac{2\pi}{\min(\Delta r_{i},\Delta y_{i})}$, which in practice leads to larger time steps.

\section{Results}
\label{sec:3}

\subsection{Model settings}
\label{subsec:1}

The parameters of our cepheid are $\mathrm{T}_{eff}=5125~\mathrm{K}$,  $\log{g}=1.97$, $\mathrm{L}=913~\mathrm{L}_{\odot}$  and  $\mathrm{M}=5~\mathrm{M}_{\odot}$. The period is about 4 days, the radius is $26~ \mathrm{Gm}$, of which the outer $11.3~\mathrm{Gm}$ were modelled. The opening angle used here is either $1^{\circ}$  or $10^{\circ}$. For the $10^{\circ}$ model the computational area has $510$ radial grid points and $800$ lateral points. For the $1^{\circ}$ model these number are $800$ and $300$, respectively. Radially the grid is stretched from cell to cell by a factor of $1.011$ and $1.07$, respectively. The radial mesh sizes of the $10^{\circ}$ model vary from $0.47 ~\mathrm{Mm}$ at the top to $124~ \mathrm{Mm}$ at the bottom.   The runs with these opening angles were started from one and the same 1D model which had been relaxed for more than 100 periods. -- The $10^{\circ}$ model is wide enough to harbour convection cells of the $\mbox{He\hspace{1pt}II}$ convection zone. Its resolution is insufficient to reasonably represent the $\mbox{H}+\mbox{He\hspace{1pt}I}$ convection zones. This purpose is achieved by the 
$1^{\circ}$ model. -- We discuss now the $\mbox{He\hspace{1pt}II}$ and the $\mbox{H}+\mbox{He\hspace{1pt}I}$ convection zone in turn.

\subsection{The $\mbox{He\hspace{1pt}II}$ convection zone}
\label{subsec:4}
After some time granted to the $\mbox{He\hspace{1pt}II}$ c.z. for proper development, ten useful periods for evaluation are at our disposal. Visualization shows a clear dependence of convection on phase. Near maximum compression the lower part of descending plumes tends to detach from the body of the c.z., and upon expansion the body reorganizes, gains strength and new plumes appear. 

\begin{figure}
\includegraphics[scale=.24]{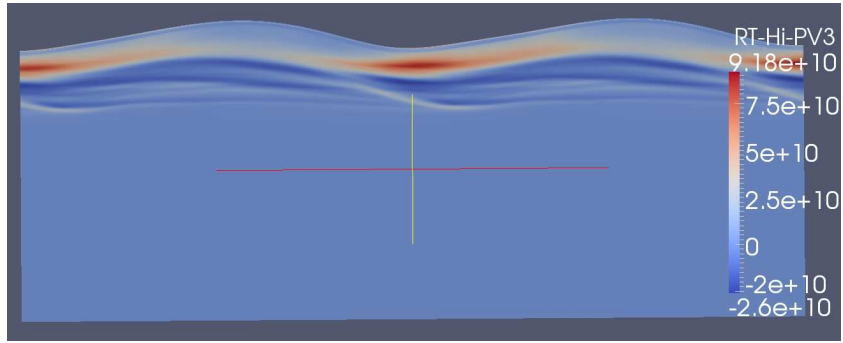}
%
%
\caption{$pdV$ work due to convection, averaged over ten periods as a function of pulsational phase. Below the convection zone proper, overshooting plumes also contribute to some extent. }
\label{fig:3}       
\end{figure}

Figure \ref{fig:3} shows phase dependence for the $pdV$ work of the $\mbox{He\hspace{1pt}II}$ zone, averaged over the ten useful periods. $pdV$ refers to the work done by convection \textit{only}, without pulsational contribution. Besides the varying strength of the work we see also some contributions due to plumes plunging into the stable region below. 

These plumes also generate noticeable gravity waves, discernible down to our lower boundary. Laterally, their extent coincides with the width of our computational domain. An even wider sector may be appropriate. They have not yet reached their statistically steady state, keeping growing.

With such models, it is also possible to test and calibrate among others convection models routinely used in modelling of radially pulsating stars. 
 
\subsection{The $\mbox{H}+\mbox{He\hspace{1pt}I}$ convection zone}
\label{subsec:5}
\begin{figure}
\sidecaption[t]
\includegraphics[scale=.23]{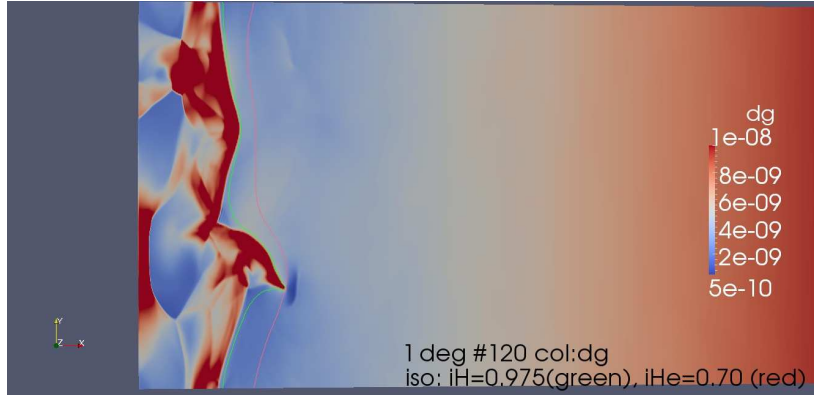}
%
%
\caption{Upper convection zone upon contraction. Top is to the left. Colours: density of gas. Isolines mark the indicated ionization degrees of hydrogen and helium.  }
\label{fig:4}       
\end{figure}

With the $1^{\circ}$ model the $\mbox{H}+\mbox{He\hspace{1pt}I}$ is reasonably resolved and, as a result,  much more vigorous than in the  $10^{\circ}$ case. Velocities (pulsation subtracted) easily get supersonic. Note the network of shocks above the hydrogen ionization front in figure \ref{fig:4}.

Frequently, shocks move upwards. In one case, this leads to the strong high-density region quite at the top. Here, our closed boundary conditions failed. We are working on implementing open boundary conditions at the top. -- Mass loss in cepheids is a topic of present discussion.  Results as the present one hint at the intriguing possibility that convectively induced shocks may be a cause. -- Generally, it should be noted that such calculations are very expensive even in 2D and by no means all cepheids are within computational reach.

\begin{acknowledgement}
Support by the Austrian Science Foundation, project P20973, is gratefully acknowledged.
\end{acknowledgement}

\end{document}